# Remote Nanodiamond Magnetometry

*Yinlan Ruan[1]\*, David A. Simpson[2], Jan Jeske[3], Heike Ebendorff-Heidepriem[1], Desmond W. M. Lau[4], Hong Ji[1], Brett C. Johnson[5,6], Takeshi Ohshima[6], Shahraam Afshar V.[1,7], Lloyd Hollenberg,[2,5], Andrew D. Greentree[4], Tanya M. Monro[1,7], and Brant C. Gibson[4]*

[1]ARC Centre of Excellence for Nanoscale BioPhotonics, Institute of Photonics and Advanced Sensing, The University of Adelaide, Adelaide, SA 5005, Australia,
[2]School of Physics, University of Melbourne, Parkville, Victoria 3010, Australia,
[3]Chemical and Quantum Physics, School of Science, RMIT University, Melbourne VIC 3001, Australia
[4]ARC Centre of Excellence for Nanoscale BioPhotonics, School of Science, RMIT University, Melbourne, VIC 3001, Australia,
[5]Centre for Quantum Computing and Communication Technology, School of Physics, University of Melbourne, Parkville, VIC 3010, Australia,
[6]Japan Atomic Energy Agency, Takasaki, Gunma 370-1292, Japan,
[7]University of South Australia, Adelaide, 5000, Australia,
\*Corresponding author: yinlan.ruan@adelaide.edu.au

Optical fibres have transformed the way people interact with the world and now permeate many areas of science. Optical fibres are traditionally thought of as insensitive to magnetic fields, however many application areas from mining to biomedicine would benefit from fibre-based remote magnetometry devices. In this work, we realise such a device by embedding nanoscale magnetic sensors into tellurite glass fibres. Remote magnetometry is performed on magnetically active defect centres in nanodiamonds embedded into the glass matrix. Standard optical magnetometry techniques are applied to initialize and detect local magnetic field changes with a measured sensitivity of 26 μT/√Hz. Our approach utilizes straight-forward optical excitation, simple focusing elements, and low power components. We demonstrate remote magnetometry by direct reporting of the magnetic ground states of nitrogen-vacancy defect centres in the optical fibres. In addition, we present and describe theoretically an all-optical technique that is ideally suited to remote fibre-based sensing. The implications of our results broaden the applications of optical fibres, which now have the potential to underpin a new generation of medical magneto-endoscopes and remote mining sensors.

## 1. Introduction

The sensing of magnetic fields is important for applications as diverse as mining exploration[1] and aircraft navigation[2]. Within the medical fields, applications such as magneto-encephalography[3] and magneto-cardiology[4] are important methodologies for sensing spatially-resolved activity in the brain and heart respectively. There are many existing magnetometers, and here we concentrate on a particular approach to optical magnetometer[5,6] integrated to an optical fiber, which has potential high sensitivity and capability to be used as a remote gradiometer .

We are focused on optical magnetometry for the sensing of fields in remote or difficult to access locations, where a small sensing volume is connected to an optical fibre for remote readout. Examples of such difficult to access locations are subterranean locations, such as within drilled holes with narrow bore for mining exploration, or within the human body. In such applications, one ideally requires a simple, robust system with a small form factor. Both SQUID (Super conducting quantum interference device) magnetometers and optical atomic magnetometers are ultrasensitive to magnetic fields[5]. Unfortunately the former needs to work in cryogenic conditions, and the latter uses dense gas as probe, therefore they can be only used inside the labs.

Here we show a new approach to optical magnetometry based on remote optically detected magnetic resonance from fluorescent nanodiamonds embedded in a tellurite glass optical fibre. Our approach uses the change in fluorescence intensity from the nitrogen-vacancy (NV) colour centre of diamond that is observed due to control of the ground state population[7,8]. We demonstrate magnetic field-induced fluorescence variation that is readout at the fibre end face, centimetres from the active nanodiamonds. Magnetometry is demonstrated using both the well-known Zeeman response[9,10], and the change in fluorescence intensity[11] from the active nanodiamonds. The latter observation offers the intriguing possibility of performing NV-based magnetometry without the use of radio-frequency driving fields, which would greatly reduce the complexity of a practical remote NV magnetometer.

The negatively-charged NV centre is the most widely studied optical defect in diamond and consists of a substitutional nitrogen atom adjacent to a carbon vacancy[7]. Its unique electron-spin properties with long-lived coherence[12] offer a solid system for applications in quantum technologies[8,13]. The NV defect has a spin-triplet ground state, which has an electronic spin that can be polarised and read out optically at room temperature[14]. The degenerate $m_s = \pm 1$ spin levels of the defect are Zeeman split in the presence of an external magnetic field with a gyromagnetic ratio of 2.8MHz/G[9]. The coherence of the NV spin has also been exploited to detect the weak fluctuating magnetic fields from nearby electronic[15-17] and nuclear spins[18,19]. The Zeeman splitting can be optically read out with greatest contrast for a single NV centre that is aligned with the unknown magnetic field. Typically this contrast is around 20%. Ensembles of NV centres reduce the contrast down to around 5%[20], although for large NV ensembles this can still provide an increased sensitivity, with the greatest demonstrated sensitivity being ~150 fT/√Hz with a millimeter scale diamond in a multi-pass configuration[21]. Alternatively, it has been proposed to use the NV centres themselves as a laser medium and this approach, termed laser-threshold magnetometry, promises f T/√Hz sensitivity[22].

The magnetic field sensitivity for NV-based systems is critically dependent on the fluorescence collection efficiency. Conventional detection approaches utilize a high numerical aperture objective positioned on top of the diamond surface, which yields a collection efficiency of around 2% due to the high index of diamond (n=2.4) leading to a total internal reflection at the diamond-air interface[23]. Such approaches are typically sensitive to mechanical noise, and limited by the size of the objective. One attractive solution to this problem is to couple the NV fluorescence into a waveguide, which reduces the overall size of the collection optics and enables facile integration with mature photonic technologies[24,25]. Several hybrid approaches have been investigated. These have focused on attaching/manipulating NV containing nanodiamonds on to the end face[26] or tapered region[27] of optical fibres. For fibre tapers coated with NDs, the fluorescence of the NVs was evanescently coupled into the fibres with maximum coupling efficiency of 1.7% achieved with a taper diameter of 360 nm[27]. The spin properties of the NV centres have been explored at the one section of the fibres[28,29], however the fragile nature of the fibre taper and lack of mechanical stability of the NDs attached to the end face devices limit the device's applicability and lifetime. In addition, the NDs were only located at one place of the fibres, thus they can be only used for a point magnetic field sensors.

In our previous work, the NDs with or without irradiated were successfully embedded in tellurite glass TZN (75TeO$_2$–15ZnO–10Na$_2$O in mol%) using a two-step glass fabrication process[30,31]. The NDs embedded in tellurite glass have been demonstrated to preserve single-photon emission from the NV centres[32], and have also demonstrated coupling of embedded NV with tellurite microspheres[33]. Our diamond glass fabrication conditions have now been optimized to enable sufficiently low loss and background fluorescence[31] so that emission from NV centres can be coupled and propagated along the fibre and detected at the fibre endface as reported in this paper. This is a significant step forward in the development of this technology and allows us to investigate the full sensing capabilities of the embedded nanodiamonds through efficient optical pumping and remote detection.

## 2. NV emission coupled to the tellurite fibre

To confirm excitation and propagation of the NV fluorescence along the tellurite fibres with 160 μm diameter, we excited the fibre from the side, which reduces background fluorescence (see Supplementary Information). We mapped the NV distribution in the fibre using an in-house built scanning confocal microscope. An area close to a fibre endface was scanned in the plane parallel to the fibre axis using a 532 nm diode laser with 7 mW power via a 100x objective (NA=0.9) attached to a nanopositioning stage. The scanned area with its size as 137 μm × 200 μm is on the plane with ~ 38 μm vertical distance to the fibre surface on the top (close to the objective) (Fig. 1a). The fluorescence image in Fig. 1c shows emission of the embedded NV centres highlighted with the dashed lines. The distortion of the emission patterns from the NV centres near the center of the fibre arises from the cylindrical shape of the fibre which acts as a lens to distort the image in one dimension. Interestingly, we can obtain a similar fluorescence image by collecting the emission from the endface and coupling the light into a multi-mode (MM) fibre as shown in Fig 1b. The endface fluorescence image shown in Fig. 1d reproduces the image features (Fig. 1c) characterised through side excitation and collection and provides a convenient readout platform for the remote detection from the embedded nanodiamonds.

To confirm the presence of NV fluorescence in each collection configuration, we compared the fluorescence spectra taken in each case. The spectra were filtered with a long-pass filter to remove the excitation light. The spectra of the bright spots numbered 1 to 7 verify NV emission, with typical spectra from two of the spots shown in Fig. 1e. It is important to note that although the fluorescence spectra were taken in two different geometries, the NV fluorescence signal is not significantly modified. Note that the larger fluorescence measured by the MM fibre is simply due to only 5% signal detected by the objective was used for spectral analysis.

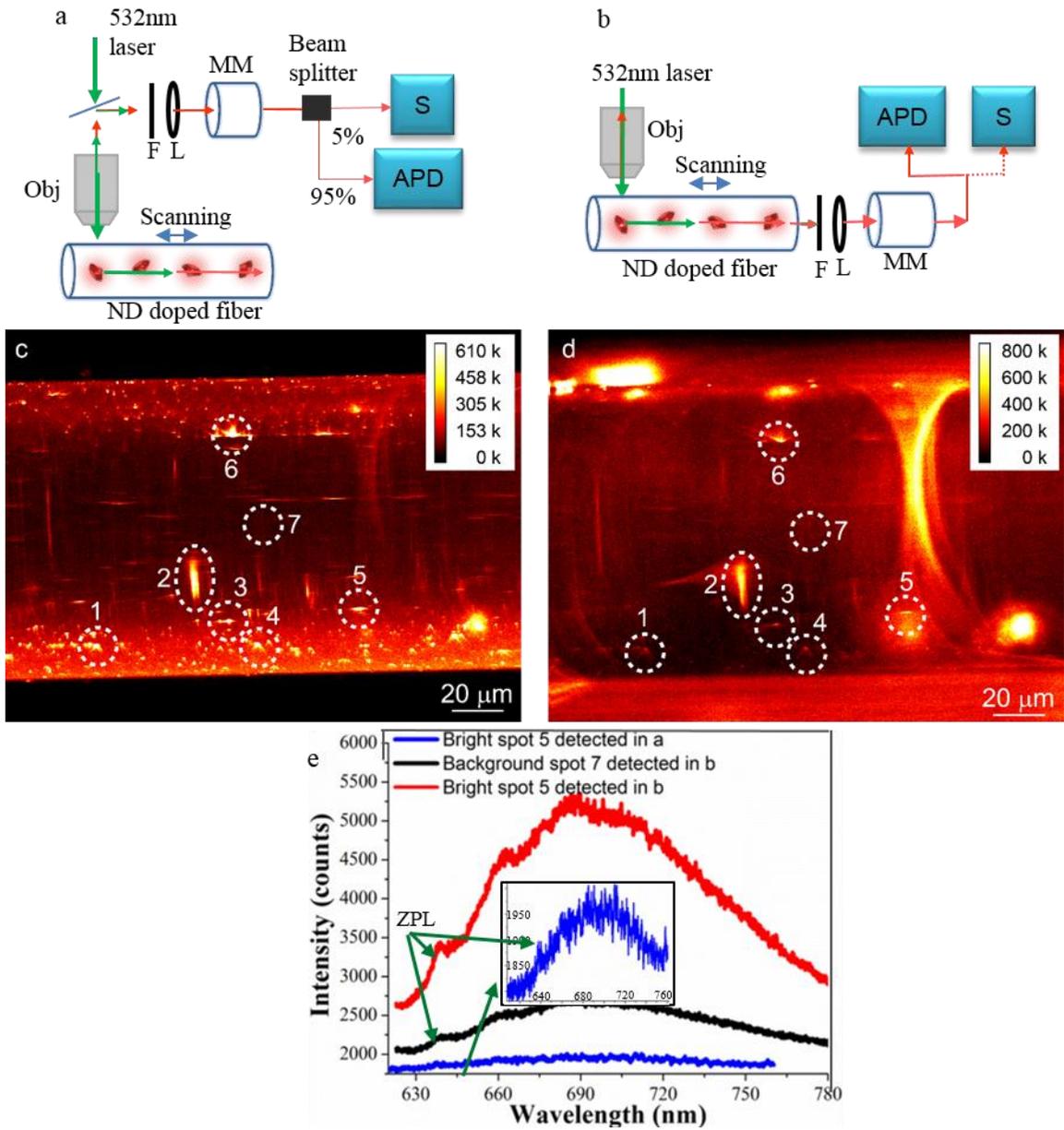

Figure 1. Mapping of the NV distribution in the nanodiamond doped tellurite fibres. a and b, Schematic side pumping setup for fluorescence mapping based on home built scanning confocal microscope. In a, The collected fluorescence signal is split into 5% and 95% for spectral and intensity measurements, respectively. In b, the output end of the MM fibre is manually connected to the spectrometer or APD for spectral and intensity measurements, respectively. F: 560nm long-pass filter, S: spectrometer, APD: avalanche photodiode detector, MM fibre: multimode fibre. Obj is100x objective (NA=0.9). c and d show the images of the scanned fibre plane parallel to the fibre axis. c was collected using the setup in a, and d was collected from b. 7 regions are numbered in c, and their corresponded positions in d are also labelled, showing these points could be viewed remotely through the fibre. e. Fluorescence spectra of Spot 5 in c (blue curve) and detected using configuration a, and Spot 5 (red curve) and 7 (black curve) in d and detected by configuration b. The excitation power was fixed as 7 mW and the integration time is 1min for all the spectra measurement.

## 3. ODMR characterization of the ND embedded tellurite fibres

Given the observation of guided fluorescence from the embedded NV centres, we now turn to the observation of ODMR (Optically Detected Magnetic Resonance) signals at room temperature using a standard collection optics and a sensitive CCD camera as shown in Fig. 2a. ODMR from NV centres in diamond is made possible due to the variation in fluorescence intensity from magnetic ground states $m_s=0$ and $m_s=\pm 1$ excited states (see Supplementary Information for further information). To investigate local response of the embedded nanodiamonds to external magnetic field, we employed a side-pumped configuration using a 200 mW 532 nm laser beam with spot size in the range of 150-200 μm. This corresponded to a light intensity (~ 600 W/cm$^2$, the same order as that of a bright LED source), about three orders magnitude higher than that we used in the previous scanning confocal microscope in Section 2. As shown in Fig. 2a, the illumination point was approximately 1 cm away from the fibre's output endface from which the guided fluorescence signal was detected by a sCMOS camera with the pump excitation removed using two long pass filters (650-750 nm). A microwave antenna was positioned just above the illumination point of the fibre. The RF power applied to the 2 mm diameter antenna ranged from 1 to 4 W. Figure 2b shows the fluorescence image of the endface of the fibre indicating a multimode pattern as expected. A permanent magnet was positioned close to the microwave antenna with controlled distance to it.

The black curve in Fig. 2c indicates a zero field ODMR at 2.876 GHz. The intensity signal is summed over the pixel imaging array and normalized to the fluorescence intensity away from resonance. The ODMR contrast of 3% is consistent with that observed from NV ensembles in the bulk diamond crystals[20]. The ODMR linewidth at zero magnetic field is 28.8 ±0.8 MHz which is broader than that observed from the ensemble NVs in a single nanodiamond. This indicates that more than one nanodiamond was probed within the excitation volume, leading to inhomogeneous broadening of the ODMR line. The sensitivity of the magnetometer to DC magnetic fields using Zeeman splitting is determined by the linewidth contrast and fluorescence count rate[34]:

$$\eta_{dc} = \frac{4\,h\delta}{3\sqrt{3} \cdot g_{NV}\mu_B R\sqrt{n}}$$

Where $\delta$ =28.8 MHz is the full width at half maximum of the ODMR peak, R = 0.015 is the ODMR contrast, $g_{NV}$ = 2.0028 is the g-factor of the NV centre[7] and $\mu_B$ is the Bohr magneton. The number of photons detected at the output end of the fibre is n = $4\times10^6$ s$^{-1}$, resulting in a calculated magnetic sensitivity of 26 μT/$\sqrt{Hz}$.

To demonstrate remote detection of external magnetic fields, our fibre-based magnetometer was placed several centimetres from a randomly oriented rare earth (Neodymium Iron Boron) permanent magnet. The ODMR signal was monitored remotely from the output end of the fibre as the applied magnetic field was increased. The Zeeman splitting of the ODMR levels induced by the external magnetic field is shown as the solid lines in Fig. 2c. In addition to the Zeeman splitting, we also observed an overall reduction in the fluorescence intensity with increasing magnetic field.

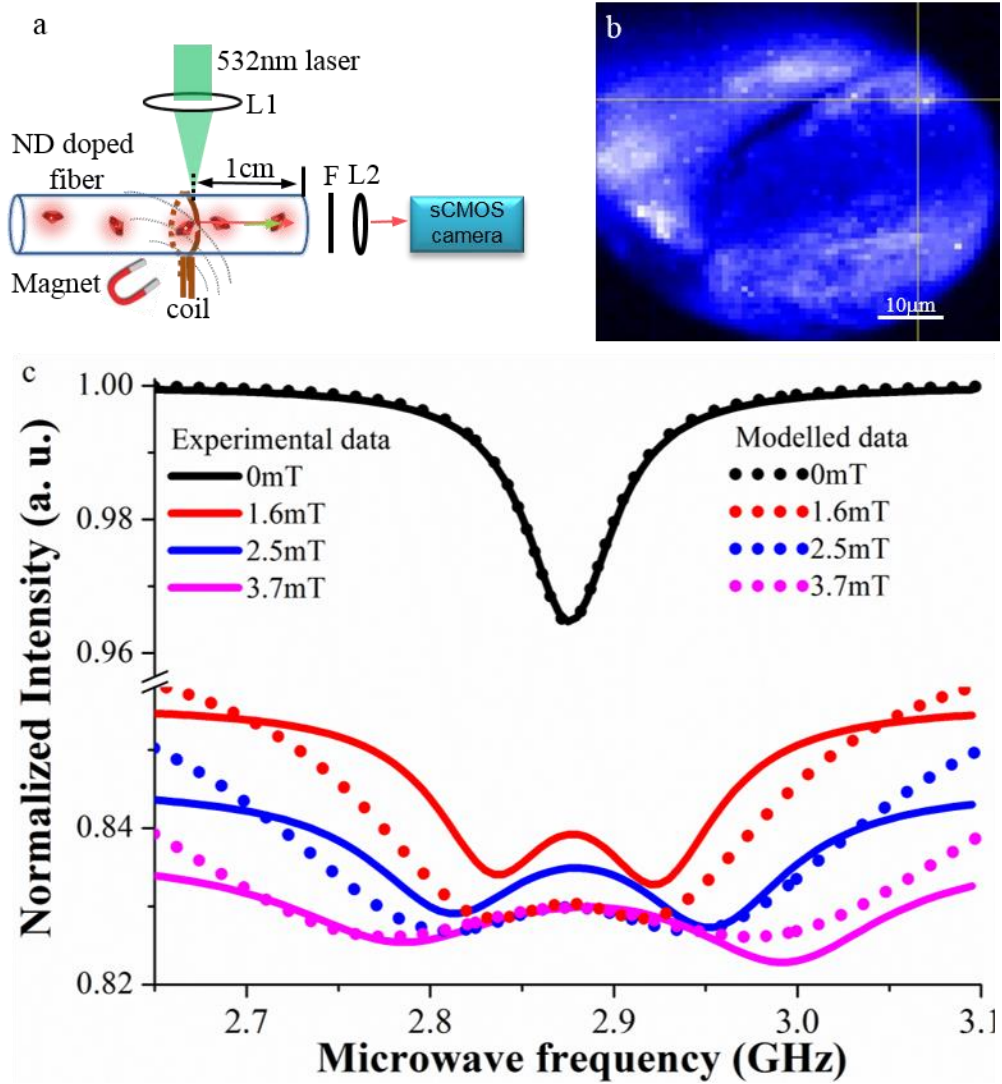

Fig. 2 ODMR characterization of the nanodiamond doped fibre. a, Experimental setup for ODMR measurement of the fibre. The fibre was excited using side pumping and the illuminated region was approximately 1 cm away from its output endface, from which the fluorescence was detected. b. Image of the fibre's output end detected by a camera, showing multimode structure. c. ODMR signals measured from the fibre's output end with increased external magnetic field (dashed lines) and the modelled ODMR signals using the same magnetic field applied in the experiments (dotted lines).

To investigate these effects we theoretically modelled the impact of the external magnetic field on the ODMR signals. We considered an ensemble of NV centres with four equally populated orientations, corresponding to the four allowed crystallographic orientations of the NV centres within a single nanodiamond particle. By following the approach in Ref[22] (see Supplementary Information), each NV centre was modelled as a seven-state system including the ground state spin triplet, the excited state triplet and a single state enabling non-spin-conserving, non-radiative transitions from the $m_s=\pm1$ excited states. We further assumed that one orientation was aligned with the remote external magnetic field. The modelled results are shown as the black dotted line in Fig. 2c. For our experimental measurements, the central ODMR resonance at 2.88GHz was split by the magnetic field as shown as solid red, blue and violet lines. We used this splitting to infer the magnetic field strength and used the same values in the modelling. Interestingly, a reduction in the overall fluorescence signal, even in the far detuned regions (<

2.7 GHz), was observed when the external magnetic fields were applied, which was in agreement with our experimental measurements shown in Fig. 2c (the solid lines). This reduction in fluorescence can be explained by the fact that spins not aligned with the external magnetic field experienced Zeeman-terms perpendicular to the local quantisation axis defined by the NV orientation and hence experienced state mixing between the 'bright' $m_s=0$ and 'darker' $m_s=\pm1$ spin states.

Although our single nanodiamond model qualitatively explains the fluorescence reduction, it does not quantitatively predict the magnitude of the fluorescence reduction, providing an overestimate to the fluorescence reduction. To match the magnitude of the fluorescence reduction, we also considered a background signal that did not depend on either RF frequency or magnetic field. We attribute this background fluorescence to residual fluorescence from the tellurite glass and scattering from NV centres that were excited by the laser and collected through the fibre but has minimal interact with the RF driving fields. We therefore treated this constant background fluorescence as a free parameter which we fitted from the data. The dash colourful lines in Fig. 2c show our modelling results after assuming the background fluorescence as 37% that of the strongest fluorescence signal (B = 0 mT). The theoretical curves are consistent with the experimental curves in Fig. 2c. Therefore our modelling results demonstrate that the background fluorescence plays a critical role on the overall performance of the device.

## 4. Discussion and Conclusion

The first fibre-based magnetometer with a magnetic sensitivity of 26 µT/√Hz and mechanical ultrastability are demonstrated using low power components. Our modelling results confirm our experimental ODMR measurements and indicate that the background fluorescence from NV centres that are not manipulated by the external microwave field have a significant impact on overall fluorescence and contrast of the ODMR peaks. The fluorescence reduction by magnetic fields away from resonance, seen in both theory and experiment, shows that there is an additional effect of the magnetic field beyond resonance splitting. This implies that all optical detection of external magnetic fields in the absence of RF fields is applicable with our device. This would largely simplify the measurement scheme although at the cost of decreased measurement precision. In this new measurement modality the fluorescence reduction would be caused only by the (relatively strong >1mT) non-aligned magnetic fields, which cause state mixing of the $m_s=0$ and $\pm1$ spin states defined by the NV orientation.

The performance of the remote fibre-based magnetometer might furthermore be improved by reducing the ND concentration in the fibre and further optimisation of the high purity raw materials for glass synthesization. This will enable a single nanodiamond to be probed within the excited region, and also lead to reduced background fluorescence of the tellurite fibre due to reduced concentration of impurities such as transition metals and rare earth ions in raw materials which can cause background fluorescence[30,35]. This would narrow the ODMR linewidth, increase contrast of the ODMR peaks and extend the magnetic field sensing range.

Since the entire length of the fibre is doped with nanodiamonds, locally pumping a small area of the fibre at different positions along the fibre length has the potential to realize a magnetic gradiometer to study the earth's magnetic field. Such devices would enable mineral material detections by embedding them into the mining holes, or realize the next generation of medical magneto-endoscopes. The beam intensity of the laser used was in the range of ~ 600 W/cm$^2$, which can be readily replaced by a LED for reduced cost and easy alignment.

## Acknowledgement

This work has been supported by ARC grants (DP120100901, DP130102494, FF0883189, FT110100225, LE100100104, FL130100119, CE140100003). T. M. acknowledges the support of an ARC Georgina Sweet Laureate Fellowship. B. C. G. acknowledges the support of an ARC Future Fellowship and the Defence Science Institute. This work was performed in part at the OptoFab node of the Australian National Fabrication Facility utilizing Commonwealth and SA State Government funding. We wish to thank Alastair Dowler at the University of Adelaide for fibre fabrication.

# Supplementary Information

## Side pumping for decreased excitation of background fluorescence

The tellurite optical fibres used here have an optical loss between 9-14 dB/m along the wavelength range from 500 nm to 800 nm. They have a concentration of the embedded irradiated NDs up to 0.7 ppm-weight and an outer diameter of 160 μm. The background fluorescence of the tellurite fibre was mainly caused by impurities included in the raw materials used for glass fabrication, and this background reduces contrast of the ODMR. Thus we characterized the effect of the glass background fluorescence in these tellurite fibres using two methods as shown in Fig. s1a and s1b. The nanodiamond embedded fibres were excited using a 532 nm continuous laser either from longitudinal direction along the fibre (Fig.s1a) or from side direction (Fig.s1b). The fluorescence output signal at the other fibre end (referred to as output end) was coupled to a spectrometer via a multimode (MM) fibre for signal analysis. A 532 nm long pass filter was positioned between the MM fibre and the spectrometer to remove light from the 532 nm pump source. Approximately 30 cm of fibre was used for this longitudinal excitation measurement. When using side pumping in Fig.s1b, the distance between the illumination point and the output end of the fibre was 8.3 cm, i.e. shorter fibre length was sampled compared to the longitudinal excitation. The normalized fluorescence spectra measured for both approaches are shown in Fig.s1c and show a characteristic $NV^-$ emission with a zero phonon line (ZPL) at 637 nm (Fig. s1c) The spectrum detected with longitudinal excitation shows an increase in the background fluorescence at short wavelengths (< 650 nm) compared to that measured by the side pumping. When the longitudinal excitation shown in Fig.s1a was used, the fluorescence from both the NV centres and tellurite glass was excited and collected along the entire length of the fibre due to the pump laser propagating along the fibre. For the side pumping in Fig. s1b, much weaker pump power was coupled to the fibre mode compared to the longitudinal excitation, resulting in largely decreased background fluorescence. The enhanced contrast of the detected NV emission to the background fluorescence by the side pumping enabled investigation of the spin properties.

## ODMR

ODMR takes advantage of the $m_s=\pm 1$ excited states being able to decay non-radiatively or radiatively, whereas the $m_s=0$ excited state only decays radiatively, thus showing brighter fluorescence. The NV centre is a seven-state system as shown in Fig. s2[8]. The radiative transitions are spin-conserving between the ground states and their excited states. By contrast, the non-radiative transition from the $m_s=\pm 1$ excited states is non-spin-conserving into the $m_s=0$ ground state (via intermediate singlet states). Due to this transition, continuous optical pumping into both the $m_s=\pm 1$ and $m_s=0$ excited states polarizes towards the $m_s=0$ states, resulting in bright photoluminescence. Application of a microwave can reduce the population of the $m_s=0$ ground state, leading to a decrease in the measured photoluminescence intensity of the NV centre. Application of a magnetic field causes Zeeman splitting of the $m_s=\pm 1$ states, with the spitting given by the NV gyromagnetic ratio 2.8MHz/G.

## Parameters for ODMR modelling

The NV centres were modelled following the approach in Ref.[22] with transition rates based on Ref.[7] (Fig. s2). The emission rate was assumed to be $\Gamma_1=(12ns)^{-1}$, the internal decay rates were $L_1=(24.9ns)^{-1}$, $L_2=(231ns)^{-1}$, $L_3=(462ns)^{-1}$, respectively. The applied microwaves are modeled as driving coherent Rabi oscillations between the ground spin triplet states with a Rabi frequency $\Omega=0.9$ MHz and an inhomogeneous dephasing rate $\Gamma_2^*=(0.01\ \mu s)^{-1}$ between the same states is assumed. The laser pumping rate was Λ=1 MHz. A laser pumping rate was assumed to be 1 MHz and the NV inhomogeneous coherence time is 0.01 μs. We note that the Rabi frequency plays a different role from that in pulsed magnetometry experiments and here the exact value is not critical, provided that the field is strong enough to create population mixing between the spin-sublevels.

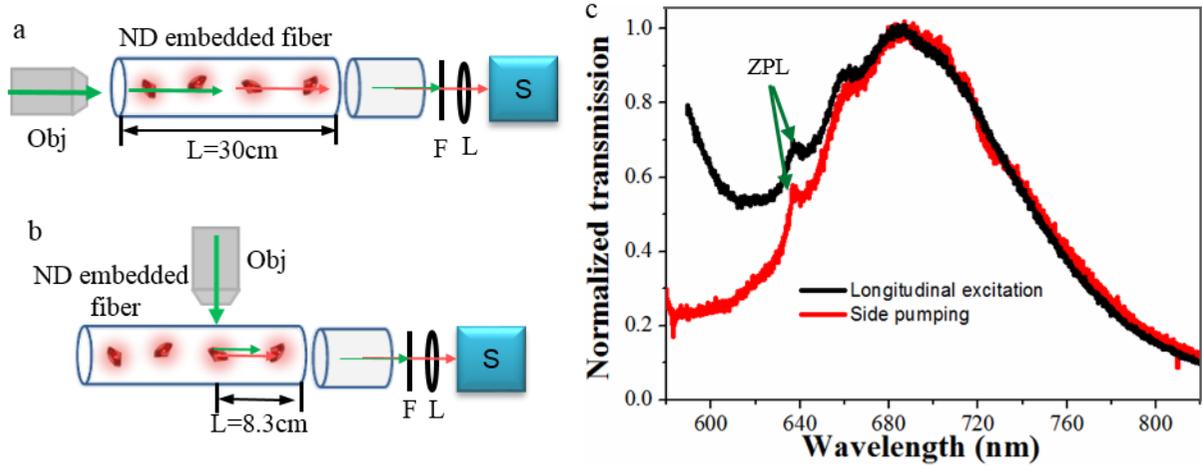

Figure s1 Fluorescence measurement of ND embedded tellurite fibres. Schematic setup for measurement of the NV fluorescence guided by the ND embedded tellurite fibres for longitudinal (a) and side (b) pumping. The propagation length of the detected NV fluorescence in (a) is 30 cm long, and is 8.3 cm in (b). A 20x objective (Obj.) with NA=0.4 was used for excitation. F: long pass filter, L: lens, S: spectrometer. c. The normalized spectra detected from the fibre output end with the black curve measured using the setup in a, and the red curve measured in b.

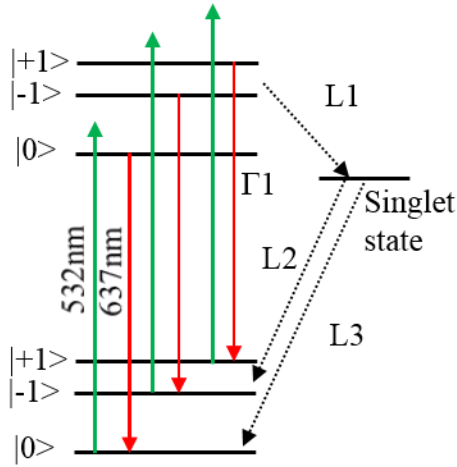

Figure s2. NV states and transitions considered in the manuscript. The solid green line is 532nm pumping laser, and the solid red line is radiative decay which enables red fluorescence at 637mnm. The dashed arrows indicate non-radiative transition.